\documentclass[%
 reprint, 
showpacs,
aps,
]{revtex4-1}

\usepackage{graphicx,amsmath,amssymb,slashed}
\usepackage{dcolumn}
\usepackage{bm}
\usepackage{hyperref}

\begin{document}

\preprint{EPJ-Plus (2016) \bf{131}: 146}

\title{ Relativistic Dynamics of the Compton Diffusion on a Bound Electron }
 
\author{Salwa Alsaleh}
\altaffiliation{salwams@ksu.edu.sa}

\affiliation{Department of Physics and Astronomy, King Saud University, Riyadh 11451, Saudi Arabia}

\date{\today}

\begin{abstract}
A covariant relativistic formalism for the electron-photon and nuclear dynamics is summarised making more accurate predictions in agreement with experiments for Compton scattering in shells with large electron binding energy. An exact solution for the Dirac equation for an electron in the nuclear Coulomb field is obtained, in order to write the relativistic dynamics for this QED process. This is a preparation for the calculation of the relativistic cross-section for Compton scattering on bound electrons; as a precision test for QED. 
\end{abstract}

\maketitle

\section{Introduction}
It is my great pleasure to write this article in honour of Monsier Directuer de recherche Frna\c{c}ois Vazeille, under the request of my student L.A Al Asfar and some of my colleagues to summarise and review the mathematical and theoretical techniques I have used in my PhD thesis for full relativistic calculation and description of Compton scattering on bound electrons \cite{my}. This description agrees completely with experimental data provided at the time I had begun writing my thesis, and up until now. It could be considered as one of the precision tests for quantum electrodynamics (QED). \\  It has been almost 30 years since I presented my calculations, and the literature still lack an exact full relativistic description of Compton diffusion on bound electrons, it is apparent that the work in my PhD dissertation needs to be reformulated in modern language and summarised in English  so the calculations and techniques become accessible to the scientific community.
Compton diffusion is an inelastic scattering process between a free, or bound electron and a photon, and it is one of most important QED processes between electrons and photons of energy less than 1 MeV. Since materialisation dominates the processes for photons of energy higher than 1 MeV.  One may consider the electron target in Compton diffusion as 'free' if its binding energy $ \epsilon$ is much smaller than the incident photon's energy $ \omega$. One then uses Klein-Nishima equation \cite{klein1929streuung} to deduce the well-known Compton relation for the incident photon's wavelength $ \Lambda$ and the emergent one's $ \Lambda'$:
\begin{equation}
\label{compton_relation}
\Lambda-\Lambda' = \Lambda_c (1-\cos \theta),
\end{equation}
with ,\begin{equation}
\equiv \Lambda_c \frac{2 \pi}{m_e},
\end{equation}
the Compton wavelength of the electron. However, the formula \eqref{compton_relation} is merely an approximation when bound electrons are considered , and fails to comply with experimental data when $ \epsilon$ is comparable to $\omega$. In particular, taking the ratio between the differential cross-sections for Compton diffusion  that was experimentally measured for the K-shell electron $ d\sigma_{exp}$ \cite{storm1970photon}; and the one calculated from Klien Nishima formula $ d\sigma_{KN}$. We find that the ratio is varying between 0.25 and 1.175, depending on the diffusion angle. Similarly for the total cross-sections ratio $\frac{ \sigma_{exp}}{\sigma_{KN}}$, that varies between 0.1 and 1. Showing a fundamental difference between free and bound electron treatment in Compton diffusion. \\
The first exact calculation of elastic scattering on bound electrons was carried by C. Fronsdal in 1966 \cite{veigele1966compton}, As for non-elastic Compton diffusion; the relativistic cross section were first approximated by \cite{gavM72} starting from non-relativistic study, then carrying out dipole approximation \cite{gavrila1972compton}. Later, inclusion of electron's spin into the calculation were made by \cite{grotch1983spin} making a quasi-relativistic calculation for the photon cross sections. More modern study made a full relativistic approximation. Other studies helped deepening the understanding of this important QED process like the studies carried by Ribbers \cite{ribberfors1975relationship} on the r\^{o}le of momentum and angular distributions, along with photon's polarisation \cite{ribberfors1975relationshipII}. A more general study  were published few months before my dissertation calculated the relativistic Compton cross sections, using Hartree-Fock central field functions \cite{PhysRevA.37.3706}.In this paper, I shall prepare for the calculation of the relativistic cress section from the covariant matrix elements. I shall start by analysing Compton scattering on bound electrons, derive a generalised Compton formula by taking into an account the nuclear recoil. Them derive the wavefunctions of the bound electrons, by analytically solving the Dirac equation exactly . Then the relativistic dynamics of the electron, photon and the nucleus are discussed at the end of this paper. 

\section{Covariant formalism }
\begin{figure}
	\label{f1}
	\centering
	\includegraphics[width= 0.9 \linewidth]{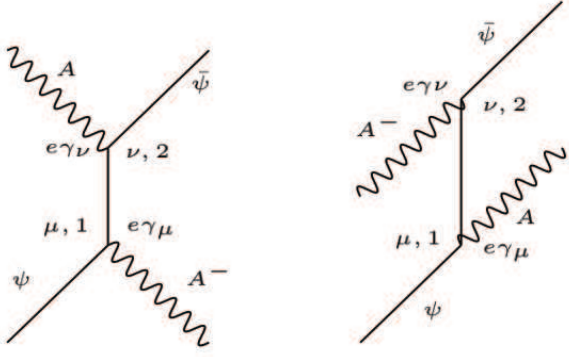} 
	\caption{Feynman diagrams for the  diffusion processes.  The diagram on the left shoes absorption first process, whilst the one on the right shows emission first process.}
\end{figure}

There are two Feynman diagrams for the Compton diffusion shown in figure \ref{f1} . The first diagram shows \emph{absorption first } process; where the incident photon is absorbed  at vertex 1 then remitted at vertex 2. The second demonstrates \emph{emission first} process, where the emergent photon is first emitted at vertex 1, then the incident photon is absorbed at vertex 2. However, these two processes are indistinguishable experimentally, hence we write the propagator of the second order process as:
\begin{equation}
\mathcal{S}= \mathcal{S}_e + \mathcal{S}_a.
\end{equation}
\begin{subequations}
	Where,
	\begin{align}
	\mathcal{S}_e = i \frac{e^2}{2} \int d^4x_1 d^4x_2 \bar{\psi}'(x_2)\slashed{A}(x_2)\mathcal{S}_c \slashed{A}'(x_1) \psi(x_1),\\
	\mathcal{S}_a = i \frac{e^2}{2} \int d^4x_1 d^4x_2 \bar{\psi}'(x_2)\slashed{A}'(x_2)\mathcal{S}_c \slashed{A}(x_1) \psi(x_1).
	\end{align}
\end{subequations}
The propagator  $ \mathcal{S}_c$ solves the equation:
\begin{equation}
\left( \slashed{\nabla}-m_0 \right) \mathcal{S}_c ( x_1-x_2) =- \delta (x_1-x_2),
\end{equation}
for the free electron, and for the bound electron case :
\begin{equation}
\left( \slashed{D}-m_0 \right) \mathcal{S}_c ( x_1,x_2) =- \delta (x_1,x_2),
\end{equation}
where $\slashed{D} \colon= \gamma^\mu \left( \partial_\mu +ie\phi_\mu \right)$, the gauge covariant derivative, and $\phi_\mu$  is an exterior field that satisfies Lorentz gauge condition $ \partial_\mu \phi_\mu =0$.  Note that the propagator no longer depend on the distance between $ x_1$ and $ x_2$, but on the points themselves.We wish to diagonalise  $\mathcal{S} (x_1,x_2)$ and write it in the form :
\begin{equation}
\mathcal{S} = \mathcal{M} \delta^4( P_1+k-P_2-k).
\end{equation}
The matrix $ \mathcal{M}$, are used to calculate the cross sections, as a substitute for Klien-Nishima formula. 
This sums the covariant formalism for the free and bound electrons in Compton Diffusion, but we shall only focus on the bound electron case from now on.
\section{Kinematic States}
In this section, I shall derive a generalised kinematic formula for Compton diffusion, by considering the Coulomb field and the recoil of the nucleus. From this section and on, I shall consider a units system where $ m_e =1$. Hence momenta have the units of $ m_e c$ and energies are written in the units of $ m_e c^2$.
\subsection{Notation}
The electron's initial and final total energies shall be respectively denoted by :
\begin{subequations}
	\begin{align}
	E_1=P_1^2= m_e+\epsilon_1 ,\\
	E_2 = P_2^2= m_e + \epsilon_2.
	\end{align}
\end{subequations}
The term $ \epsilon_1 = T_1 +V_1$, corresponds to the electron's initial kinetic energy $ T_1$, and its binding energy with the nucleus $V_1$. The final energy term $ \epsilon_2$ is only given by the final kinetic energy, as the final electron is considered to be free, viz $ \epsilon_2 \simeq T_2$. \\ As for the photon's initial and final 3-momenta, they are denoted by $\vec{k}$ and $\vec{k}'$, where the energy is given by $ \omega = |\vec{k}|$ and $ \omega'= |\vec{k}'|$ . And the 4-momenta satisfy the 'null-like' condition :
\begin{equation*}
k^\mu k_\mu = k'^\mu k'_\mu =0 ,
\end{equation*}
since the initial and the final photon states are real. 
Finally, for the nucleus states, the initial state of the nucleus of a mass $M$ is considered to be static, thus the only energy it has is the one corresponding to its mass. Whilst in the final state the nucleus will acquire a recoil kinetic energy $q_r$. The 4-momenta for the nucleus are written as:
\begin{subequations}
	\begin{align}
	Q= ( 0, M ), \\
	Q^\prime= (\vec{q}_r, M).
	\end{align}
\end{subequations}
\subsection{ The Generalised Formula} 
Since the 4-momenta are discussed above, we can set up the diffusion in the laboratory system of reference (LAB) as figure \ref{refsys} shows. The initial scattering occurs in the $ x,z $ plane, the incident photon approaches from the positive $z$ direction. From the figure one can write the following geometric relation:
\begin{figure}
	\label{refsys}
	\centering
	\includegraphics[width= 0.9 \linewidth]{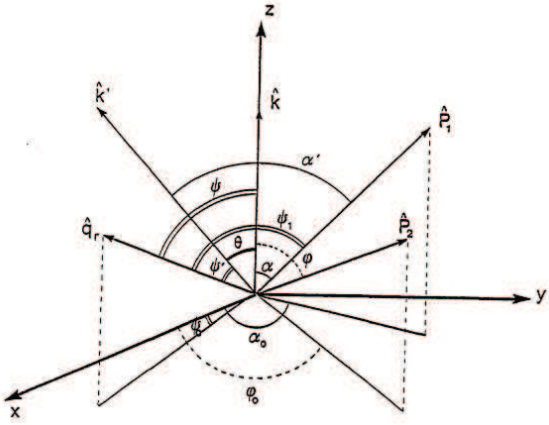} 
	\caption{ Laboratory system of reference, the nucleus is in the centre and the scattering is happening in the $(x,z)$ plane. The incident photon comes from $ z+$ direction.} 
\end{figure}

\begin{equation}
\cos \alpha' = \cos \alpha \; \cos \theta + \; + \sin \alpha \, \cos \alpha_0 \, \sin \theta .
\end{equation}
Then one may determine the scattering angle of the electron with respect to the nucleus by $\alpha$; that takes values $ [0, 2\pi]$, and the initial angle for the electron is $ \alpha_0 $ , that takes the values $ [ 0, \pi] $.  The angels $ \psi$ and $ \psi '$ determine the position of the nucleus in the LAB system. they are related to other angels by the relations:

\begin{align}
\cos \psi_1& = \sin \alpha \, \cos \alpha_0 \; \sin \psi \; \sin \psi_0 \; + \sin \alpha \; \sin \alpha_0 \; \nonumber \\ \qquad & \times \sin \psi \; \sin \psi_0 + \cos \alpha \; \cos \psi  \\
\cos \psi'& = \sin \theta \; \sin \psi \; \cos \theta \; + \cos \psi.
\end{align}

From these geometric relations, and basic conservation laws. A generalised formula for Compton diffusion can be written in the simple form :
\begin{equation}
\Lambda' -\Lambda =  2 \pi \rho .
\end{equation}
The factor $\rho$ is written as:
\begin{equation}
\rho = \frac{1}{2\pi}  \frac{\sum\limits_{i=0}^{K} \Delta \Lambda_i}{\sum\limits_{i=0}^{K} R_i}
\label{master}
\end{equation}
The couple $ \{ \Delta \Lambda_i  R_i\}$ determine the approximation in mind, depending on the value of $K$ as follows:
\begin{itemize}
	\item{ K=0 . The Classical Compton formula \cite{compton}; for free and stationary electron.}
	\begin{align}
	\Delta \Lambda_0 =& 2 \pi \left( 1- \cos \theta \right),\nonumber \\ 
	R_0 =& 1 .
	\end{align}
	\item{ K=1, Dumond's formula \cite{dumond1933linear} for free but moving electron. The Doppler effect correction is included  \eqref{master}.}
	\begin{align}
	\Delta \Lambda_1 =& - \left(P_1 \lambda / E_1) \right) \left(\cos \alpha - \cos \alpha ' \right),\nonumber \\ 
	R_1=& - \left(P_1 / E_1)\right) \; \cos \alpha .
	\end{align}
	\item {K=2, Correction for electron's binding with the nucleus is obtained . Matching the results of Ross and Kirkpatrick \cite{ross1934constant}.}
	\begin{align}
	\Delta \Lambda_2=& - \left(\Lambda^2/ 4 \pi E_1) \right) \left[(P^2_2-E^2_2)-(P^2_1-E^2_1)\right],\nonumber \\
	R_2=&- \frac{ \Delta \Lambda_2}{\Lambda}.
	\end{align}
	\item{ K=3, Corrections for moving and bound electron are obtained; the recoil of the nucleus is taken into an account. Matching the results of Veigele \textit{et al.} \cite{veigele1966compton}.}
	\begin{align}
	\Delta \Lambda_3 =& -\frac{ \Lambda q_r }{E_1} \; \left( \cos \psi - \cos \psi'\right) \; - \frac{\Lambda^2 P_1 q_r}{2\pi E_1} \cos \psi_1, \nonumber\\
	R_3=& \frac{q_r}{2 \pi E_1} \left(P_1 \Lambda  \cos \psi_1 - 2 \pi \cos \psi \right).
	\end{align}
\end{itemize}
As we can see, the most general case is the one obtained from taking $K=3$. In some cases the nuclear recoil plays an important r\^{o}le in the scattering process that cannot be neglected. Particularly when the recoil 3-momentum $ Q$ reaches its maximal value; this can be deduced from the geometry of the diffusion:
\begin{equation}
Q_{Max} = P_1+\omega + P_2+ \omega'.
\end{equation} 
The last parameters of the formula \eqref{master} can be rewritten in this case as :
\begin{align}
\Delta \Lambda_3 =& -\frac{ Q_M }{E_1} \; \left(2+ \Lambda P_2 / 2 \pi \right) , \nonumber \\
R_3=& \frac{Q_M}{2 \pi E_1} \left(P_1 \Lambda  +2\pi \right).
\label{maxrec}
\end{align}

The formula \eqref{master} can be used to calculate the final energies for the photons and electron:
\begin{align}
\omega'=& \frac{\omega}{1+\rho \omega}, \nonumber \\ 
E_2=&\frac{\rho \omega^2}{1+\rho \omega }+ ( 1+ \epsilon_1).
\end{align}
However, both formulae depend on the nuclear recoil. In order to overcome this dependence, one can consider the limiting cases.Considering the cases when the photon is completely absorbed ($ \omega'=  0$), or when it bounces off without loosing any energy to the electron i.e $ E_2 = 1$, and $ \omega'$ takes its maximal value. We obtain, a formula for the maximal/minimal values of $\omega$ :
\begin{widetext}
\begin{equation}
\omega_{max/min}= \omega \times \frac{ A \left[ \left(B-A\cos\theta \right)^2+\sin^2\theta \left(A^2-P_1^2\right)\right]^{1/2}\pm P_1\left(B-A\cos\theta\right)}{B\left[ \left(A-B\cos\theta \right)^2+\sin^2\theta \left(B^2-P_1^2\right)\right]^{1/2}\mp P_1\left(A-B\cos\theta\right)}.
\end{equation}
\end{widetext}
Where:
\begin{equation}
A= E_1- \frac{1}{2\omega} \left(P_1^2-E_1^2+1\right),
\end{equation}
and
\begin{equation}
B= E_1+\omega\left(1-\cos \theta\right).
\end{equation}
The formula above is a correction to Compton's formula for bound electrons that allows the calculation for the final 4-momenta for electron in the limiting cases specified above. \\
The full relativistic dynamics of electron resembled by its initial 4- momentum is however more difficult to compute, it requires the calculation of the wavefunctions for the electron using Dirac equation. This shall be the task for the next section.
\section{Bound Electron's Relativistic Wavefunction}
Adopting the notation in the previous section, Dirac equation for the bound electron takes the explicit form :
\begin{gather}
\label{dirac}
\left(-i \vec{\alpha} \vec{\nabla} + \beta +V-E\right) \psi(\vec{r}) = 0,  \nonumber \\
\vec{\alpha}= i\beta \vec{\gamma} ,\nonumber \\
\beta= \gamma^4.
\end{gather}
Where $ ( \vec{\gamma} , \gamma^4)$ \footnote{The 'particle physicist convention  is used, with $4$ as the time index label, and metric signature of 'mostly minuses'.}, is the set of Dirac gamma matrices, with spinor indices suppressed. \\The Dirac spinor $\psi(\vec{r})$ can written as -in van der Warden representation-:
\begin{equation}
\psi(\vec{r}) = \begin{pmatrix}
v(\vec{r})\\
u(\vec{r})
\end{pmatrix}.
\end{equation}
Thus \eqref{dirac} reads -eliminating the spinor $u(\vec{r})$ - :
\begin{widetext}
\begin{equation}
\nabla^2 v(\vec{r}) + \left\lbrace \left[ (E-V)^2 -1+(E-V+1) \right] \left[ \vec{\sigma}\vec{\nabla} ( E-V+1)^{-1} \right]  \vec{\sigma}\vec{\nabla} \right\rbrace  v(\vec{r}) =0 .
\label{dirac2}
\end{equation}
\end{widetext}
Where $\vec{ \sigma} = ( \sigma ^1 , \sigma ^2 , \sigma ^3)$ are the set of Pauli spin matrices.\\
Clearly \eqref{dirac2} is separable, the solution takes the form :
\begin{equation}
v(\vec{r}) = g(r) \chi (\phi, \theta).
\end{equation}
The first function is the radial function $g(r)$, while the second resembles the angular part $\chi(\phi,\theta)$.  The latter function contains the coupling between the orbital $L$ and spin $S$ angular momenta for the electron, it can be written as :
\begin{equation}
\chi ^\mu _{jl} = \sum_{s} C ( l \frac{1}{2}j ;\mu-s,s) \; Y^{\mu-s}_{l} \; \chi^s .
\label{angular_wavefunction}
\end{equation}
where:
\begin{itemize}
	\item{$ j, l$ are the total and orbital angular momenta quantum numbers, respectively}
	\item{$\mu$ is the eigenvalue for the z-component of the angular momentum $ \hat{J}_z$.}
	\item{$ \chi^s$ , is Pauli spinor corresponding the the spin eigenvalues $ s = \pm \frac{1}{2}$. Such that\\ $ \chi ^{+1/2} =$ $1 \choose 0$ and $ \chi ^{-1/2} =$$ 0 \choose 1$.}
	\item{$Y^{\mu-s}_{l}$ , the normalized spherical harmonics}
	\item{$C ( l \frac{1}{2}j ;\mu-s,s)$ , the corresponding Clebsch-Gordan coefficients with respect to the quantum numbers mentioned above.}
\end{itemize} 
One can express the wavefunction in \eqref{dirac} as:
\begin{equation}
\psi = (\vec{r},E) = \binom {1}{-i ( E-V+1)^{-1}\sigma_r(\partial_r+\frac{\kappa+1}{r})} \times g_{jl}\;  \chi ^\mu _{jl}
\end{equation}
This expression is deduced from the expression from the orbital angular momentum operator representation in the spherical-polar coordinates:
\begin{align}
\vec{\sigma} \cdot\vec{\nabla} = \sigma _r ( \partial_r - \frac{1}{r}  \vec{\sigma}\cdot	\vec{L} ), \nonumber \\
\vec{L} = i\vec{r} \wedge \vec{\nabla},
\end{align}
the number $ \kappa $ resembles  the angular quantum number that shall be discussed below. \\ 
Now, the spinor $ u(\vec{r}) $ can be recovered,  and expressed in analogous manner to $ v(\vec{r})$, the radial function  $f_{jl}(r)$ is introduced corresponding to the spinor $ u(\vec{r}) $ - similar to $g_{jl}(r)$. The radial parts of Dirac equation then becomes :
\begin{align}
\left( \partial_r + \frac{\kappa+1}{r}\right)g(r)_{jl}- ( E-V+1) f(r)_{jl} =0 ,\\
\left( \partial_r - \frac{\kappa-1}{r}\right)f(r)_{jl}+( E-V-1) g(r)_{jl} =0 .
\end{align}
The wavefunction now takes the form:
\begin{equation}
\psi^\mu _{jl} = \binom{g_{jl}}{-if_{jl}} \chi ^\mu _{jl}
\label{wavefunction1}
\end{equation}
Which will be useful for later calculations of the dynamics of Compton scattering.\\ 
Now we turn to define the angular quantum number $ \kappa$. This quantum number can be defined as the eigenvalue of the operator $ - ( \vec{\sigma}\cdot \vec{L}+1)$ with the angular functions $ \chi ^\mu _{jl} $, viz:
\begin{equation}
- ( \vec{\sigma}\cdot \vec{L}+1) \chi ^\mu _{jl} =  \kappa \chi ^\mu _{jl}.
\end{equation}
Hence, $\kappa$ depends of the three quantum numbers $ j,l$ and $s$ :
\begin{equation}
\label{kappa}
\kappa  = l (l+1)- (j+ \frac{1}{2})^2,
\end{equation}
noting it is either larger than $l$ or less than $ -l-1$. Taking both positive and negative values, but never zero. \\ It is then tempting to express the angular momenta quantum numbers in terms of $ \kappa$ guided by the expression for it in \eqref{kappa}. Moreover, writing the eigenfunctions $\chi ^\mu _{jl}$ and $ \psi^\mu _{jl}$ , more conveniently, in terms of $\kappa$. That is
\[ 
\chi ^\mu _{jl} \rightarrow \chi ^\mu _\kappa,
\]
and so on. 
We can also define the action of the operator $ \sigma_r \colon= \vec{\sigma }\cdot \vec{r}$ on $\chi ^\mu _\kappa$ as :
\begin{equation}
\sigma_r \chi ^\mu _\kappa = - \chi ^\mu _\kappa.
\label{sigmar}
\end{equation}
Finally we are ready to write the final form of the wavefunction, noting that the solution \textit{only} depends on the quantum numbers $ \mu$ and $ \kappa$; thus rewritings eq \eqref{wavefunction1}  using the previous result from \eqref{sigmar} as: 
\begin{equation}
\psi^\mu _{\kappa} = \binom{g_{\kappa} \chi ^\mu _{\kappa}}{if_{\kappa} \chi ^\mu _{\kappa}} 
\label{wavefunction2}
\end{equation}
\subsection{Radial wavefunctions}
\subsubsection{Continuous wavefunctions}
In order to calculate the radial wavefunctions that appear in \eqref{wavefunction2}, one needs to find the potential. In this study, the potential is the central Coulomb potential of the nuclei $ V= \frac{- \alpha Z }{r}$, with $\alpha$ being the fine-structure constant, and $ Z$ is the atomic number. The argument followed here is similar to \cite{rose1961relativistic}, but with the potential $V$ taken into an account. \\ Taking $ E \geq 1$, the spectrum of the electron's wavefunctions should be \textit{continuous}. The radial wavefunctions therefore is expressed as :

\begin{multline}
rf_\kappa = \frac{ i(E-1)^{1/2} ( 2Pr)^\gamma e^{\pi y/2} | \Gamma ( \gamma+iy)|}{2( \pi P )^{1/2}\Gamma ( 2 \gamma+1)} \\ \times \left[e^{-iP+i \eta} \left(\gamma +iy \right) F (\gamma+1+iy, 2\gamma+1, 2iPr) + cc.\right],
\end{multline}
\begin{multline}
rg_\kappa = \frac{ i(E+1)^{1/2} ( 2Pr)^\gamma e^{\pi y/2} | \Gamma ( \gamma+iy)|}{2( \pi P )^{1/2}\Gamma ( 2 \gamma+1)} \\ \times \left[e^{-iP+i \eta} \left(\gamma +iy \right) F (\gamma+1+iy, 2\gamma+1, 2iPr) + cc.\right].
\end{multline}
Where, $F (a, b, c)$ and $ \Gamma(z)$ are the hypergeometric and gamma functions respectively. The factors appearing the expressions above are defined as the following:
\begin{align}
P &= (E^2-1)^{1/2}, \nonumber \\
\gamma &=( \kappa^2 - \zeta^2)^{1/2}, \nonumber \\
y&= \zeta \frac{E}{P}, \nonumber \\
\zeta &= \alpha Z .
\end{align}
The factor $ e^{ i\eta} $ is a normalisation factor and it is found to be :
\begin{equation}
e^{i \eta} = S \frac{ ( \kappa -\gamma )P+ \zeta(E-1)}{\left[2(E-1)(E\kappa-\gamma)(\kappa-\gamma)\right]^{1/2}}.
\end{equation}
Defining the factor $S$ in similar fashion to \cite{overbo1968exact}:
\begin{equation}
S= \frac{1}{2}\left[1+\frac{EZ}{|EZ|}+\frac{\kappa}{|\kappa|}\left(\frac{EZ}{|EZ|}-1 \right)\right],
\end{equation}
thus $S = \pm1 $ , depending on $ E,Z$ and $\kappa$.\\
\subsubsection{ Discrete wavefunctions}
For electron's energy $ -1<E<+1$, the radial wavefunctions should be discrete. The take the following general form (following the lead of \cite{rose1961relativistic}) :
\begin{widetext}
\begin{equation}
f_\kappa = - \frac{2^{1/2}\lambda^{5/2}}{\Gamma (2\gamma+1)}\left[\frac{\Gamma( 2\gamma+n'+1)(1-E)}{n'!\zeta (\zeta-\lambda \kappa)}\right]^{1/2} (2\lambda r)^{\gamma -1} e^{-\lambda r} \left[n'F(-n'+1,2\gamma+1, 2\lambda r)-\left(\kappa -\frac{\zeta}{\lambda}\right)F(-n',2\gamma+1, 2\lambda r)\right],
\label{fwave}
\end{equation}
\begin{equation}
g_\kappa =  \frac{2^{1/2}\lambda^{5/2}}{\Gamma (2\gamma+1)}\left[\frac{\Gamma( 2\gamma+n'+1)(1+E)}{n'!\zeta (\zeta-\lambda \kappa)}\right]^{1/2} (2\lambda r)^{\gamma -1} e^{-\lambda r} \left[-n'F(-n'+1,2\gamma+1, 2\lambda r)-\left(\kappa -\frac{\zeta}{\lambda}\right)F(-n',2\gamma+1, 2\lambda r)\right].
\label{gwave}
\end{equation}
\end{widetext}

With $ E = 1+\epsilon$, $\epsilon$ being the electron's relativistic binding energy, $n '= n - |\kappa|$, with $ n$ being the principal quantum number, and $ \lambda = \sqrt{1-E^2}$. The solutions  $ f_\kappa$ and $ g_\kappa$, contain the hypergeometric function with negative 'entries', hence they can be related to Laguerre polynomials $\mathcal{L}^{\alpha}_{\beta}$. One may express the solution in terms of Laguerre polynomials and Gamma function, in order to make numerical calculations more efficient, first defining :
\begin{equation}
N= 2^{1/2}\lambda^{5/2} \left[\frac{\Gamma( 2\gamma+n'+1)}{n'!\zeta (\zeta-\lambda \kappa)}\right]^{1/2}.
\end{equation} 
Then the radial wavefunctions are written as :
\begin{widetext}
\begin{equation}
f_\kappa =  -N \sqrt{(1-E)} (2\lambda r)^{\gamma -1} e^{-\lambda r}   \left[n'\frac{\Gamma( n')}{\Gamma(2\gamma+n')}\mathcal{L}^{2\gamma}_{n'-1}  ( 2 \lambda r)  -\left(\kappa -\frac{\zeta}{\lambda} \right)\frac{\Gamma(n'+1)}{\Gamma(2\gamma+n')+1}\mathcal{L}^{2\gamma}_{n'}  ( 2 \lambda r)\right]
\end{equation}
\begin{equation}
g_\kappa = - N \sqrt{(1+E)} (2\lambda r)^{\gamma -1} e^{-\lambda r}  \left[n'\frac{\Gamma( n')}{\Gamma(2\gamma+n')}\mathcal{L}^{2\gamma}_{n'-1}  ( 2 \lambda r) +\left(\kappa -\frac{\zeta}{\lambda} \right)\frac{\Gamma(n'+1)}{\Gamma(2\gamma+n')+1}\mathcal{L}^{2\gamma}_{n'}  ( 2 \lambda r)\right]
\end{equation}
\end{widetext}

\begin{figure}
	\label{radwavefuncition2s}
	\centering
	\includegraphics[width= 0.9 \linewidth]{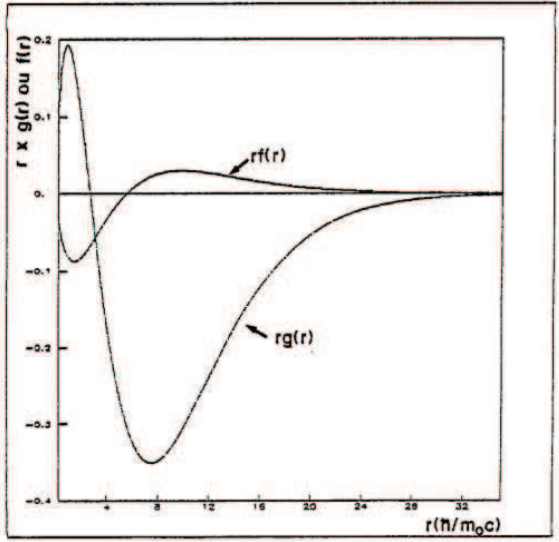}
	\caption{The radial wavefunctions multiplied by $r$ for the state $2S_{1/2}$ for $Z= 82$. }
\end{figure}

\begin{figure}
	\label{radwavefuncition2p}
	\centering
	\includegraphics[width= 0.9 \linewidth]{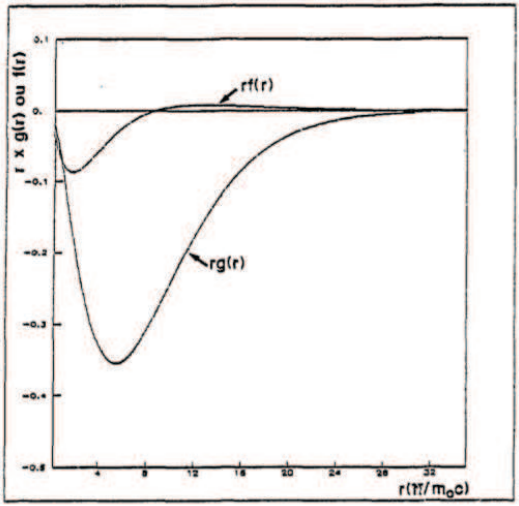}
	\caption{The radial wavefunctions multiplied by $r$ for the state $2P_{1/2}$ for $Z= 82$. }
\end{figure}

The ratio $ f_\kappa / g_\kappa$, satisfies the limit  as $ r\rightarrow \infty $, and $ n'=0$:
\begin{equation}
\frac{f_\kappa}{g_\kappa} \sim - \sqrt{\frac{1-E}{1+E}}
\end{equation}
One can find a 'simple' expression for the discrete radial functions, for $n$ taking values from $1$ to $6$. Notice for shells with these principle quantum numbers, we have $ 2n-1$ states per shell, determined fully by $ n'$ and $ \kappa$.  The simple form of the wavefucntions is expressed as the following:
\begin{subequations}
	\begin{align}
	f_\kappa&= -\mathcal{N} \sqrt{( 1-E)} e^{-\lambda r} r^{\gamma-1}P_{n'}(r), \\
	g_\kappa&=\mathcal{N} \sqrt{( 1+E)} e^{-\lambda r} r^{\gamma-1}Q_{n'}(r).
	\label{simple}
	\end{align}
\end{subequations}
Where $P_{n'}(r)$ and $Q_{n'}(r)$, are polynomials of $r$ of the $n'$th order :
\begin{align}
P_{n'}(r) \colon= \sum\limits_{m=0}^{n'} a_m r^m ,\nonumber \\
Q_{n'}(r) \colon= \sum\limits_{m=0}^{n'} c_m r^m,
\label{ac}
\end{align}
with
\begin{align}
a_0 = ( n'-\kappa)E+\gamma+n' ,\nonumber \\
c_0 = - ( n'+\kappa)E+\gamma+n'.
\end{align}
Moreover, for $ m \neq 0$, we can use the special functions representation of $ f_\kappa$ and $ g_\kappa$ to find the m th coefficients $ a_m $ and $c_m$ :
\begin{align}
a_m ( m \neq 0) = B_m \lambda^m \frac{ (a_0 -mE)}{(2 \gamma +m)(2\gamma+m-1) \dots( 2\gamma+1)}, \nonumber \\
c_m ( m \neq 0) = B_m \lambda^m \frac{ (c_0+mE)}{(2 \gamma +m)(2\gamma+m-1) \dots( 2\gamma+1)},
\end{align}
\begin{equation}
B_m = \frac{ (-2)^m \Gamma( n'+1)}{m!\Gamma(n'+1-m)}.
\end{equation}
We may also define :
\begin{gather} 
\mathcal{N}  =\frac{1}{2 \Gamma(2 \gamma+1)}\cdot \frac{(2 E\zeta)^{\gamma+1/2}}{( \gamma+n')^{\gamma+1}} \cdot \left[\frac{\Gamma(2\gamma+n'+1)}{n'!(n'!+\gamma-\kappa E)}\right]^{1/2}, \nonumber \\
E= \frac{n'+\gamma}{[n^2-2n'(|\kappa|-\gamma )]^{1/2}} \\
\lambda= \frac{E \cdot \zeta}{\gamma+n'}. \nonumber 
\end{gather}
Notice that from this expansion of the radial wavefunctions, it is clear that they depend on $ E, r$ and $ \kappa$, but not on $ \mu$. 
The formulae above are used to calculate the radial wavefunction for the state $ 2S_{1/2}$ for $ Z =82$ in the figure \ref{radwavefuncition2s}, and for state $ 2P_{1/2}$ for the same element figure \ref{radwavefuncition2p} , as an example. We observe that the radial wavefunction for the $ f_\kappa $ tends to be smaller and smaller as the shell gets farther and farther from the nuclear Coulomb field. On the converse, it becomes considerable for shells closer to the nucleus.
We are ready now to write the total wavefunction $ \psi^\mu _\kappa$ for the Dirac field around the nucleus. This wavefunction will be important to describe the scattering process when we shall discuss the propagators.  We shall used the formula \eqref{angular_wavefunction} to describe the angular and spin part of the total wavefunction, the formulas \eqref{fwave} and \eqref{gwave} or their 'simple'  form \eqref{simple} to describe the radial part. Noting that for Compton diffusion for bound electrons in the shells close to the nucleus, we ought to consider both radial wavefunctions $ f_\kappa$ and $ g_\kappa$, as both of them play a r\^{o}le as we saw. \\
\section{Dynamic Electron States}
Since we have the wavefunction for the bound electron  $ \psi^\mu _\kappa$, it is possible to study the dynamics of the electron in parallel to what was established in the kinematics section. The full dynamics of the electron - before the diffusion process- is described by its initial  momentum $ P_1$ and its binding energy $ \epsilon_1$. A full relativistic calculation for these is possible since the relativistic wavefunction is found. 
\subsection{ Relativistic binding energy}
The relativistic binding energy for the electron is given by the formula  \cite{rose1961relativistic} :
\begin{equation}
\epsilon_1 = \left[ 1+\left( \frac{ \zeta}{n-|\kappa_1|+\gamma}\right)^2 \right],
\end{equation}
keeping the same parameters from previous section. 
\subsection{Relativistic momentum}
In order to calculate the momentum of the electron, we need fist to calculate the expectation value of its square, then take the square root of that result, viz:
\[
P= \sqrt{\langle P^2\rangle}
\]
The subscript in $ P_1$ shall be dropped throughout the calculations of this section, as this argument is valid for \textit{any} dynamical state for the electron.  The value $\langle P^2\rangle$ is calculated from:
\begin{equation}
\label{expect}
\langle P^2\rangle = \int d^4 x  \bar{\psi^\mu _\kappa} P ^\dagger P \psi^\mu _\kappa.
\end{equation}
The kinetic energy is given by \cite{rose1961relativistic}:
\begin{equation}
\vec{\alpha} \cdot \vec{P} = -i \vec{\alpha} \cdot \vec{r} \partial_r+i\frac{\vec{\alpha} \cdot \vec{r}}{r} \vec{\sigma}\cdot \vec{L}
\label{moment}
\end{equation}
In fact, the product  $ (\vec{\alpha} \cdot \vec{P}) ^\dagger (\vec{\alpha} \cdot \vec{P})$ yields the operator $ P^2$. Thus substituting \eqref{moment} in \eqref{expect}, and by the orthonormal property of the angular wavefunctions in \eqref{angular_wavefunction}, we have:
\begin{equation}
\langle P^2\rangle = \int dr\, g^*(r) a(r, \kappa)g(r) r^2 + \int dr\, f^*(r) a(r, \kappa)f(r) r^2.
\label{expected2}
\end{equation}
Where :
\begin{equation}
a(r,\kappa ) = \frac{\partial^2}{\partial r² }+\frac{(\kappa+1)^2}{r^2}- \left[r^{-2} -2r^{-1} \frac{\partial}{\partial r}\right]( \kappa+1),
\end{equation}
is the position representation for the operator $P^2$ obtained from \eqref{moment}.
To calculate the integrals in \eqref{expected2} we use the formula for the radial wavefunctions in \eqref{simple}, we obtain after lengthy but straightforward calculation :
\begin{widetext}
\begin{equation}
\langle P^2\rangle = \mathcal{N}^2 \sum_{l=0}^{n'} \sum_{m=0}^{n'} \frac{\Gamma( 2\gamma +m+l-1)}{(2 \lambda)^{ 2\gamma +m+l-1}} \left[ H_0+ \frac{ 2\gamma +m+l-1}{2\lambda} H_1 +\frac{( 2\gamma +m+l-1)( 2\gamma +m+l)}{(2 \lambda)^2}H_2 \right].
\label{momentum}
\end{equation}
\end{widetext}
With:
\begin{equation}
H_i = (1-E)a_m F_i+(1+E)c_m \cdot G_i ,
\end{equation}
\begin{center}
	for $i = 0,1,2 . $ , and :
\end{center}
\begin{align}
F_0 &= B_0+ \kappa (\kappa -1)a_m+2(1-\kappa)A_0 \nonumber \\
F_1&= B_1+2(1-\kappa)A_1 \\
F_2 &= B_2  \nonumber
\end{align}
\begin{align}
G_0 &= D_0+ \kappa (\kappa +1)c_m+2(1+\kappa)C_0 \nonumber \\
G_1&= D_1+2(1+\kappa)C_1 \\
G_2 &= D_2  \nonumber
\end{align}
\begin{align}
A_0 &= ( \gamma+m-1)a_m \nonumber \\
A_1&= -\lambda a_m
\end{align}
\begin{align}
B_0 &= ( \gamma+m-1)( \gamma+m-2)a_m \nonumber \\
B_1&= -2\lambda ( \gamma+m-1)a_m\\
B_2 &= -\lambda^2a_m \nonumber
\end{align}
\begin{align}
C_0 &= ( \gamma+m-1)c_m \nonumber \\
C_1&= -\lambda c_m
\end{align}
\begin{align}
D_0 &= ( \gamma+m-1)( \gamma+m-2)c_m \nonumber \\
B_1&= -2\lambda ( \gamma+m-1)c_m \\
B_2 &= -\lambda^2c_m \nonumber
\end{align}
We notice from the expression\eqref{momentum} that the momentum depends on the 'effective' atomic number $ Z_{eff} = Z-\mathcal{Q}$ , with $ \mathcal{Q}$ being the charge density of the other electrons. In other words, one cannot ignore the interaction between the electron of study and other electrons in the other shells \cite{gavM72}.
We can use the previous formulae to make  numerical calculations for $ \epsilon$ and $P$  and compare them with the non-relativistic ones as in tables \ref{t1}  and \ref{t2} for the element $ Z= 82$ (lead).
\begin{table}
	\label{t1}

	\centering
	\begin{tabular}{|l|c|c|c|c|}
		
		\hline
		Shell & K & L$_{I}$& L$_{II}$ & L$_{III}$ \\ 
		\hline
		$\kappa$ &  $-1$& $-1$& $1$ & $-2$ \\
		\hline
		$\epsilon$ ( non relativistic)&  $0.17903$& $0.04475$&$0.04475$&$0.04475$ \\ 
		\hline
		$\epsilon$ (relativistic)&$0.19879$  &$0.05100$ &$0.05100$ &$0.04581$ \\
		\hline
		
	\end{tabular} 
		\caption{Calculation of the electron's binding energy in the shells $K$ , $L_{I} , L_{II}$ and $ L_{III}$ in units of $ m_0c^2$ for $ Z= 82$.}
\end{table}
\begin{table}
	\label{t2}

	\centering
	\begin{tabular}{|l|c|c|c|c|}
		
		\hline
		Shell  & K & L$_{I}$& L$_{II}$ & L$_{III}$ \\ 
		\hline
		$\beta$ &  $0.5986$& $0.2993$& $0.2993$ & $0.2993$ \\
		\hline
		$P$ ( non relativistic)&  $0.7472$& $0.3136$&$0.03136$&$0.03136$ \\ 
		\hline
		$P$ (relativistic) &$0.9438$  &$0.2945$ &$0.1639$ &$0.3186$ \\
		\hline
	\end{tabular} 
		\caption{Calculation of the electron's momenta in the shells $K$ , $L_{I} , L_{II}$ and $ L_{III}$ in units of $ m_0c$ for $ Z= 82$.}
\end{table}
From the initial binding energy and momentum, one can use the formula \ref{master} to know the final state of the electron after the diffusion. 
\section{The Nuclear Dynamics}
\begin{figure}
	\label{noya}
	\centering
	\includegraphics[width=0.9 \linewidth]{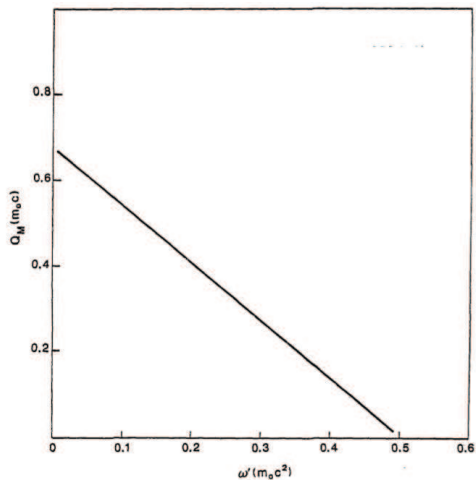}
	\caption{Estimation of the maximal nuclear recoil as a function of $ \omega'$ for $\omega = 662 keV$ and $ Z= 32$.}
\end{figure}

The maximal nuclear recoil is when the final electron is state is a free electron, viz the electron is liberated from the atom due to the diffusion process. We are interested in calculating the nucleus recoil in this case. The nuclear momentum $Q$ and kinetic energy $ T_r$ are related by conservation of 4-momenta by:
\begin{equation}
T_r = \sqrt{( Q^2+M^2)} -M .
\label{recoil}
\end{equation}
Using  the equation above and equation \eqref{maxrec} we arrive to the equation that $ Q_{M}$ must satisfy :
\begin{widetext}
\begin{equation}
Q_{M}^4-4M^2\left[ 1+ \frac{E_1+\omega-\omega}{M} \right] Q_M ^2 +8M^2(P_1+\omega +\omega')Q_M+4M^2 [(E_1+\omega -\omega')^2- ( P_1+\omega +\omega')^2-1] =0
\end{equation}
\end{widetext}
Since the initial states for the electron can determined from the previous section, and the photon's final energy is a measurable quantity, the only quantity left is $ \omega$ to be that can be determined a priori within the allowed ranges.
The figure \eqref{noya} demonstrates the nuclear recoil momentum for Compton diffusion on K-shell electron in element $Z= 32$ (Germanium) with incident photon's energy $ \omega = 662$keV 
Since $ Q_M$ depends on the kinematics of the electron and the photon described by the generalised Compton formula \eqref{master}, and at which degree of approximation described by $K$. 

\section{Concluding Remarks}		
A generalised Compton formula is derived for moving bound electrons, from a geometrical argument. This formula allows the calculation for the final electron's energies and momenta, and predicts the final photon energy and nucleus recoil provided the incident photon's energy is known. The covariant formalism for the diffusion process in quantum electrodynamics is discussed for bound electrons, where the binding energy from the Coulomb potential is no longer negligible.This, along with finding the electron's relativistic wavefunction prepares for the calculation of the full relativistic cross sections, that are supposed to be more aligned with experimental results than previous calculations. The relativistic wavefunctions are obtained from analytical solution of the Dirac equation for electron in (strong) field. It is evident from this solution that for internal shells (particularly $K$ and $L$), the target electron possess 'large' and 'small' radial wavefunctions corresponding to its interaction with the virtual particles created from the (strong) electromagnetic field present in the vicinity of the nucleus. More effects are shown to take part in the diffusion process for the target electron,from the study of its dynamics. For example its interaction with other electrons in the other shells, and the nuclear recoil after the scattering process. 
This paper provided a detailed discussion and calculation for this important QED process, and prepares for further exact calculation of a (generalised) relativistic cross section that are made in other paper.\\
\section*{Acknowledgement}
	This research project was supported by a grant from the '\textit{ Research Center of the Female Scientific and Medical Colleges} ', Deanship of Scientific Research, King Saud University.

\bibliographystyle{spphys}       
\bibliography{ref}   

\end{document}